\documentclass[aps,prd,superscriptaddress,eqsecnum,amsfonts,showpacs,epsf]{revtex4}
\usepackage{epsfig}

\newcommand{\be}{\begin{equation}}
\newcommand{\ee}{\end{equation}}
\newcommand{\bea}{\begin{eqnarray}}
\newcommand{\eea}{\end{eqnarray}}

\newcommand{\nn}{\nonumber}

\begin{document}

%\preprint{ \parbox{1.5in}{\leftline{hep-th/0111433}}}

\title{Hadronic vacuum polarization in  $e^{+}e^{-}\to \mu^{+}\mu^{-}$ process bellow 3 GeV}

\author{V. \v{S}auli}

\email{sauli@ujf.cas.cz}
\affiliation{Department of Theoretical Physics, Institute of Nuclear Physics Rez near Prague, CAS, Czech Republic  }

\begin{abstract}
The interference effect between leptonic radiative corrections and hadronic polarization functions is calculated
via optical theorem for $\mu-$pair productions. It is achieved within the use of the last data for 
$\sigma_h(e^+e^-\rightarrow hadrons)$
from SND,CMD,CMD-2,BABAR and BESSIII experimental Collaborations. The paper takes into account specific experimental conditions 
and the result is compared with KLOE experiment for $\mu^{-}\mu^{+}$ production at $\phi$ meson energy. Small, but non-negligible tension between theory and experiment is discussed.
  
\end{abstract}

\maketitle

%%%%%%%%%%%%%%%%%%%%%%%%%%%%%%%%%%%%%%%%%%%%%%%%%%%%%%%%%%%%%%%%%%%%%%%%%%%%%%%%%%%%%

%
\section{Introduction}

Comparisons between theory and experiment are used to test Standard Theory for decades. 
Folded in Monte Carlo codes for Bha-Bha and $e^{+}e^{-}\to \gamma\gamma $ processes it also serves for measurement 
of luminosity in modern colliding beam facilities used by  experimental 
collaborations around the world.  For less then percentage accurate measurement the studies
require consistent account of leptonic as well as hadronic virtual corrections. The hadronic contribution to photon  vacuum 
polarization function plays particularly important role, since it is the main source of uncertainties in theoretical calculation
of muon anomalous magnetic moment $a_{\mu}$. The last precise measurement of $a_{\mu}$ , together with the last decades data 
for electrohadron production, leads to an evidence of tension between Standard Theory and the experiments  \cite{amu1,amu2}. 
Similar confrontation of the theoretical technique with the experimental accuracy is offered by long time known \cite{BERKOM1976} interference effect between leptonic and hadronic vacuum polarization functions in close vicinity of  narrow resonances: $\omega$ and $\phi$ as well as heavier quarkonia the $J/\Psi,\Psi$ and  $\Upsilon$'s. There, the hadronic vacuum polarization is enhanced several orders of magnitude above hadronic background, which together with precise measurement of muon production $e^{+}e^{-}\to \mu^{+}\mu^{-}$ at $\phi$ energy by the SND and the KLOE detector, allows to  test the Standard  theoretical assumptions in a similar manner as for   $a_{\mu}$. 

The purpose of this paper is twofold. In the second place, the calculation of hadronic polarization function  $\Pi_{h}(s)$ up to $3 GeV$ is  performed within the use of the last data on exclusive meson production from BESSIII, BABAR,CMD,CMD-2 and SND measurements.
We collect data for total cross sections of the processes $e^+e^->\pi\pi,\pi\pi\pi,K^+K^-,K_LK_s$ and  made an interpolating fit of all cross sections separately. BABAR and BESSIII collaborations used ISR method and cover very large energy region.  We have found it is hard or impossible  to decouple resonances separately as their parameters are strongly correlated due  to the presence of their neighbors.  Thus in order to get rid of numerical obstacles, we have  generated quasidata  and  calculate iteratively $\Pi_{h}$ through the optical theorem by principal value integrations. 
As the first and the main  purpose, we implement obtained hadronic polarization into the formula for theoretical cross sections 
for $e^{+}e{-}\to \mu^{+}\mu^{-}$ and compare with the available data and predict otherwise. We discuss the errors. Within these days experimental errors on $\sigma_h$, we can see small tension between theory and measured muon-production cross section.

\section{Theory of $e^{+}e^{-}\rightarrow \mu^{+}\mu^{-}$ for KLOE}

With the income of the first  direct evidence \cite{AUGU1973} for hadronic vacuum polarization in electron-positron annihilation in 1973, the importance of radiative and soft photon correction has been recognized \cite{BERKOM1976}. 
The unique interference effect has been obtained within the help of Breit-Wigner formula already in 1963 in 
 \cite{Gato1963}. The estimate was  based on the following replacement:
\be \label{old}
e^2\rightarrow e^2\left(1-\frac{3m_V\Gamma_{ll}\alpha^{-1}}{m^2_V-s+im\Gamma}\right) \, ,
\ee
which was made in the leptonic cross section. Within the Eq. (\ref{old}), the $\phi$ resonance effect has been studied  quite recently \cite{ACH2000,ACH2001,AMBRO2005} and used for the  identification of  leptonic widths.
In fact validity of formula (\ref{old}) rise and fall with the size of the resonance term. The larger the interference effect is, the worse the first order expansion (\ref{old}) of the full polarization function $\Pi$ necessary becomes.

 In what follow we compare calculated integrated cross section 
 $\sigma(e^{+}e^{-}\to \mu^{+}\mu^{-}) $ with the high precision measurement obtained by KLOE detector.
The  detector provided three points with the best (two promile) accuracy ever achieved, 
which is enough to  test the Standard Theory in a  strict way typical for quantities $a_{\mu}$ or $\alpha(M_Z)$. 
For this purpose  we integrate analytically  the differential cross section formulas as defined
 in \cite{ARBU1997}. The integrated cross section can be written in the following way
\be \label{etomu}
\sigma(s)=\frac{4\pi C_t}{\left|1-\Pi(s)\right|^2}
\left[\sigma_A(s)\left(2-\beta_{\mu}^2(1-\frac{C_t^2}{3})\right)+\sigma_B(s)\right]
 \, ,
\ee
where $C_t$ stand for $\cos(\theta_{min})$ with $ \theta_{min}=50^0$ ($\theta_{max}=140^o$), which is KLOE  experimental cut on polar scattering angle between $\mu^{-}$ and $e^{-}$ particles and $\beta_{\mu}=\sqrt{1-4m^2_{\mu}/s}$.
Further  we have defined  the function $\sigma_A(s)$  such that it is given by the  part of the differential cross section,
that has an angular dependence  identical to the Born cross section. It uniquely defines the rest,  which explicitly reads   
\be
\sigma_B(s)=-\frac{\alpha^3}{4\pi s}
(1-\beta_{\mu}^2)\ln{\frac{1+\beta_{\mu}}{1-\beta_{\mu}}} \, .
\ee
Thus the main term  $\sigma_A(s)$, listed completely in \cite{ARBU1997}, collects all  leading logs of Dirac and Pauli form factors and the known soft photon contributions for which we take $ln\frac{\Delta \epsilon}{\epsilon}=0.05$ (15 MeV cut on c.m.s. soft photon energy at $\phi$ peak).  
Let us mention that the the both $\sigma_A$ as well as $\sigma_B$ are slowly changing  real valued functions and do not play any role in the observed interference effect. Rather, they define the norm of the integrated cross section.

All interesting interference effect stems from the sum of leptonic $\Pi_l$ and hadronic polarization function. Both, they are complex and their sum constitutes the polarization function 
\be
\Pi(s)=\Pi_l(s)+\Pi_h(s) \,\, ,
\ee
and completes the running QED coupling:
\be \label{alfa}
\alpha(s)=\frac{\alpha}{1-\Pi(s)} \, ,
\ee
with $\alpha=\alpha(0)=1/137.0359991390$ understood. 

The imaginary part of $\Pi$ is given by cross section $\sigma_h$ by virtue of the dispersion relation
\cite{CABGAT1961,EIDJEg1995}:
\be \label{muf}
\Pi_h(s)=\frac{s}{4\pi^2\alpha}\int_{4m_{\pi}^2}^{\infty}d\omega \frac{\sigma_h(\omega) \left[\frac{\alpha}{\alpha(\omega)}\right]^2}
{\omega-s+i\epsilon}\, .
\ee

The Optical Theorem, which is a must of the theory, together with the additional but  usual analytical assumptions are  milestones on the path of derivation of the dispersion relation (\ref{muf}). Thus, these are not only errors of experimental data on $\sigma_h (\Pi_h)$, which limits the precision of the final leptonic cross section, but there can be also some systematics if the expression for $\Pi_h$ slightly differs from the equation (\ref{muf}). Hence one is testing Standard Theory, rather then the Standard Model.
Note also the Eq. (\ref{muf}) produces also a real  poles  in limiting case of  ideally narrow resonances.

Purely QED contributions are  well known from perturbation theory.
Since, there are mistakes in the formula  in Ref. \cite{ARBU1997}, I present leptonic contributions into the vacuum polarization function here:
\be
\Pi_l(s)=\frac{\alpha}{\pi}\Pi_1(s)+\left(\frac{\alpha}{\pi}\right)^2\Pi_{2e}(s) \, \, ,
\ee
where one loop contribution is
\bea \label{lepton}
\Pi_1(s)&=&\Pi_e(s)+\Pi_{\mu}(s)+\Pi_{\tau}(s)\, \, ;
\nn \\
\Pi_f(s)&=&-5/9-x_f/3+f(x_f)\,\, ; f=e,\mu,\tau \, ; 
\nn \\
f(x_f)&=&\frac{\beta_f}{6}(2+x_f) 
\left(\ln{\frac{1+\beta_{f}}{1-\beta_f}} -i\pi\right)\Theta(1-x_f)
+\frac{\beta_f}{3}(2+x_f)  arctg\left({\frac{1}{\beta_f}}\right)\Theta(x_f-1)  \, \, ,
\eea
where $\beta_f=\sqrt{|1-x_f|}$ and $x_f=4m^2_f/s$.
Also the leading logarithmic term stemming from the second order
\be
\Pi_{2e}(s)=\frac{1}{4}ln(\frac{s}{m_e^2}-i\pi)+\zeta(3)-5/24 \, .
\ee
is taken into account. Recall in advance here, that in addition to leptons, the one  loop formula for heavy quarks, has an usual extra factor $\alpha\rightarrow \alpha N_c e_q^2$.

Separation of  individual meson peaks by using Breit-Wigner formula was used to  determine $\Pi_h$ consequently \cite{EIDJEg1995}, which was  standard strategy well justified by  large errors for exclusive hadroproduction data of last millennium. To get more precise result, one should avoid doubts due to the resonance extraction from the form factors. In fact, even quite narrow resonances like $\phi, \Psi..$ are coupled to their neighbors and their interference should be properly counted for. At these days, in order to calculate $a_{\mu}$ and  to perform the integration over the dispersion relation kernels, the method requires more careful account of  experimental data  \cite{DAVIER2011,BENA2014}. Also for us, since we are dealing with principal value integration, the interpolation, averaging and integration procedure  is the state of art of the method.
Recall, the use of  experimental data too much straightforwardly would lead to numerical noise and lost of accuracy required. Thus before of all, the well established interpolating fit to the existing experimental data for $\sigma_h$ for each exclusive channel is found separately. It consequently  allows to use   arbitrarily large number of generated  quasidata points, which  makes systematic errors from principal value integration in (\ref{muf})  immaterial. Also, the fit of experimental errors (as a function of s) is required for each combinations of measurements. The quality of obtained fits is the keystone for comparison between theory and experiments.  Achieving small $\chi^2$ for all fits, then this is solely the fitted error function $\epsilon_h(s)$, which most easily will tell us how the experimental errors of $\sigma_h$ propagate into the function $\Pi_h$ and muon pair production cross section.

This millennium published data are used as they have been collected by experimental collaborations. 
 SND,CMD,CMD2, KLOE detectors were operating near $\phi$ and $\omega$ mesons, while others like BABAR and BESS covered the full low energy region. I use four main exclusive channels of $\sigma_h$:$\pi\pi$, $K^{+},K^{-}$, $K_LK_S$ and $\pi\pi\pi$  measured in various experiments \cite{BESIII,BABAR2012,SND2006} for $\pi\pi$,
, further  \cite{SND2002,SND2003,babar2004} for $3\pi$ and  \cite{CMD,CMD2} for $K^{+},K^{-}$, $K_LK_S$, $3\pi$ and  \cite{ACH2007}for  $K_LK_S$. Since covering 5 GeV energy region, very important data  for physics of $\phi$ meson are taken from  2013 precision measurement of the cross section $e^{+}e^{-}\to K^{+}K^{-}$ presented in Supplemental Material  of the paper 
\cite{LEES2013}. We neglect the higher multiplicity  processes like $4\pi$, $2K+2\pi$ \cite{BELLE2009}, etc., noting here, that they contribute by amount of several $nb$ in restricted maximum region and their contributions can be neglected.  The existing cross sections fits (see for instance \cite{BABAR2012,LEES2013}) based on various phenomenological models for  were explored, modified and refitted. Our aim is to get good interpolation of $\sigma_h$, rather then  to "improve" Breit-Wigner parameters of existing PDG data.
The details of these fits will be available in planned future publication.
 We have minimized $\chi^2$ for strongly correlated parameters -masses, widths and  phases  of  resonances- as they typically appear, but having no intention to improve PDG, we freely  take these parameters  as an independent ones for every process (thus for instance $m_{\phi}$ in KK modes can differ about 0.4 MeV from the one in $3\pi$ mode).

\begin{figure}
\centerline{\epsfig{figure=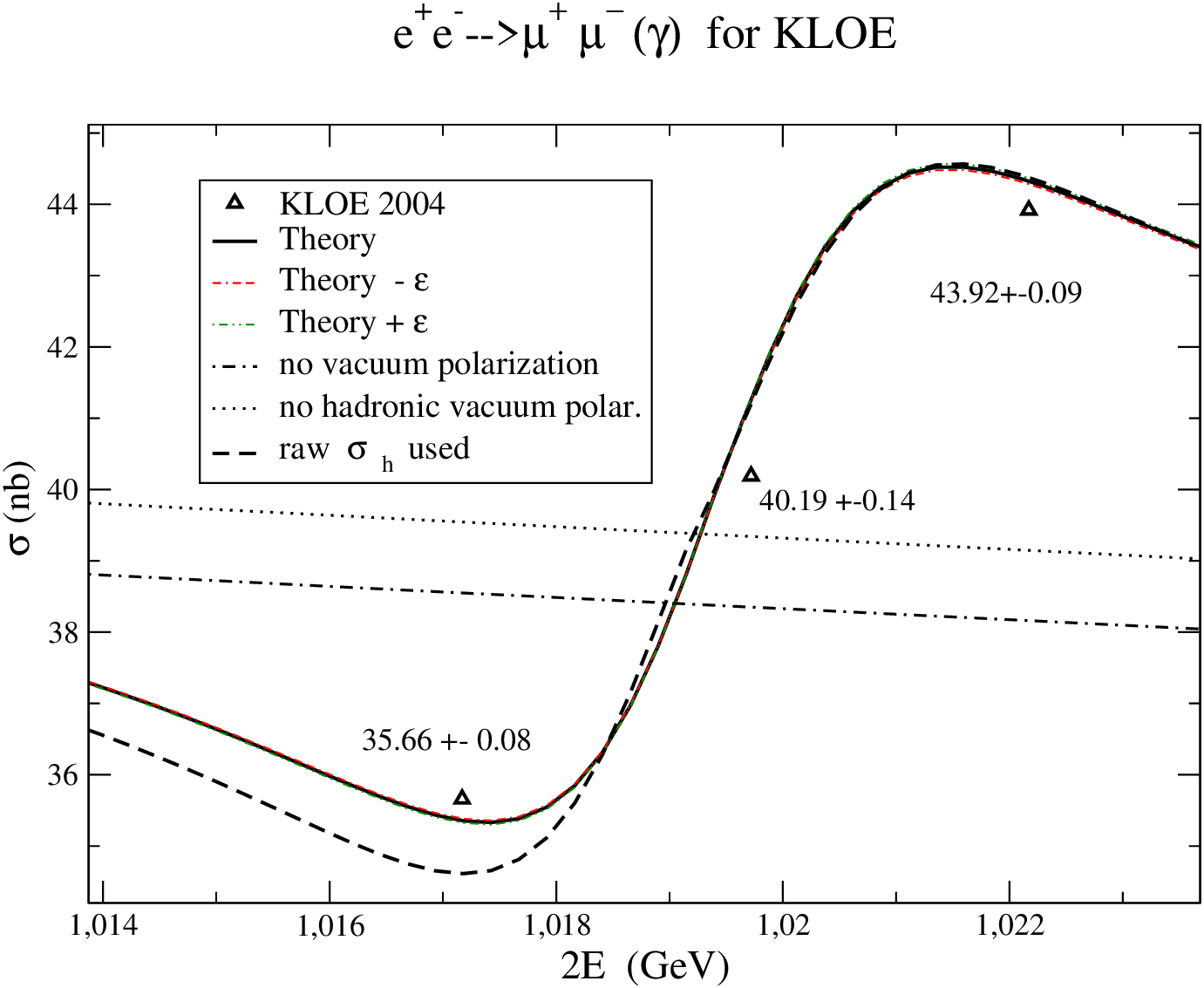,width=12truecm,height=10truecm,angle=0}}
\caption[caption]{Muon pair cross sestion, comparison between theory and experiment as described in the text.} \label{figure1}
\end{figure}

The result is shown in the Fig. \ref{figure1} where we also compare with approximation which neglects the running 
coupling in the numerator of rhs. (\ref{muf}). The KLOE data points are shown with statistical errors explicitly in the graphs. 
One can see that the final result (represented by solid thick line) with correct -vacuum subtracted- cross sections $\sigma_0$ taken into account, (see the formula (\ref{muf}) ) differs remarkably from the case when one uses 'zero iteration' raw cross section $\sigma_h(s)$ instead of $\sigma^0_h(s)$ (dashed line). 
Two thin solid lines were obtained with shifted cross sections by the statistical error of $\sigma_h$.  It offers rough estimate of the propagation of experimental  errors. However,  it does not represent the maximal error which could be find by more sophisticated Monte-Carlo method (considering all allowed area between the two curves $\sigma_h(s)\pm\epsilon(s)$). Doing this explicitly is challenging   numerical task remaining  for future.  We also compare to the cross section
with  hadronic polarization switched off (dotted line) and to the case when polarization function is switch off completely (dashed dotted line).

\section{Discussion of the results}

\begin{figure}
\centerline{\epsfig{figure=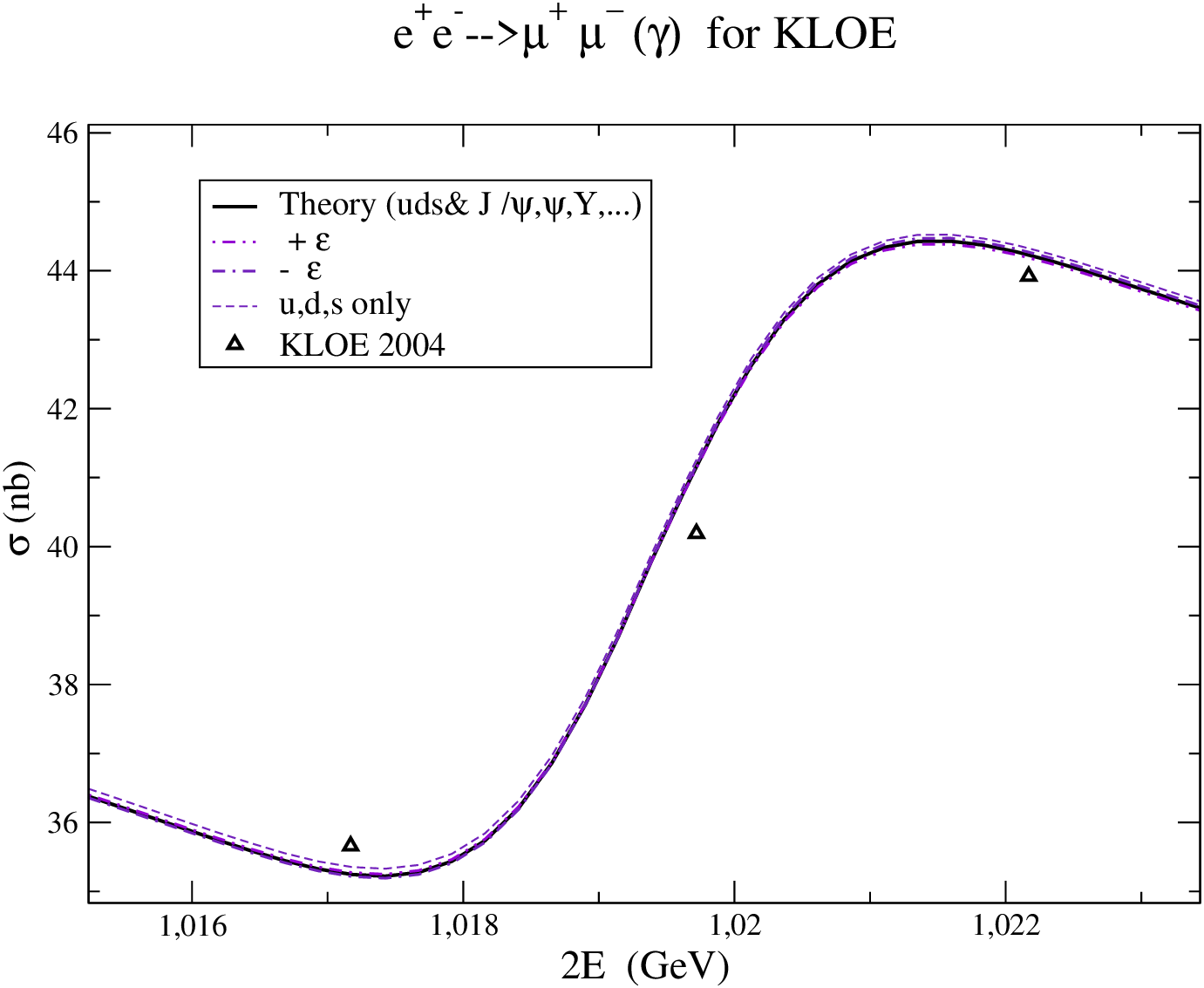,width=12truecm,height=10truecm,angle=0}}
\caption[caption]{Muon pair cross section, comparison between theory and experiment as described in the text.} \label{figure2}
\end{figure}

The observed interference effect is roughly reproduced by the Standard Theory dispersion relation. One gets 95 percentage  agreement between theory and experiment when one considers the difference (or slope) of KLOE measured points. Naively, the observed  theory-experimental tension is comparable  with the one in  the case of anomalous magnetic moment.
However in our case, the effect is theoretically  overestimated and  we get five (four) standard deviations difference when comparing   theoretical and experimental values at the left (the right) shoulder of zig-zag interference structure. 
Note also, the theory/experiment difference is negative in the first point, while it is positive in the second one, leaving quite  small space for the systematic errors.  Actually, this is the genuine  asymmetric shape of $\phi$ interference, which allows to ignore relatively larger theory/experimental discrepancy for the center KLOE point.
 Assuming unrevealed systematic error and allowing a possible small shift in a beam energy determination as well as in  the cross section  norm, the sum of  differences between theory and experiments  would not vanish at all. Of course,  systematical error not completely determined and known, can slightly lower the tension between theory and experiment. Stressed here for clarity, we do not fit the theoretical curve to experimental points, as this is not consistent with small error for $\sigma_h$.

In addition I offer simple arguments, showing that 
the observed discrepancy is not over- but rather under-estimated. It will not be relaxed if one ads the off resonance missing contributions in $\sigma_h$. 
First of all, let us stress that bellow $\phi$ peak, the all exclusive process are measured very precisely.  Thus for instance
there is a little change when taking $\sigma_h\pm\epsilon_h$, with
the  error function $\epsilon_h(s)$ artificially increased such that it  exceeds the actual experimental errors several times in  the whole region of  $\rho$ meson peak. Only the change  as large as ten percentage of the total $\rho$ peak can move the theoretical curve to the experimental value of the left KLOE point (and it  moves the remaining two points upwards).
The similar can be stated for the effect coming from  the changes in  $\sigma_h$  above $\phi$ meson peak. 
Again, small missing and omitted parts of hadronic cross section, which are reasonably above the peak,  makes the total discrepancy almost intact.  Let us show the effect of larger changes, e.g. the one which is given by the  most prominent part of missing contribution to $\sigma_h$, that we have ignored up to now. It is a charm-anticharm structure with its main contributor to $R_h=(\sigma_h/sigma(\mu\mu)$
 - the $J/\psi$ meson.
In order to estimate the effect at the scale $s\simeq m_{\phi}$ we have included $J/\psi$ and all other well established vector (V) charmonia and bottomonia in a usual simplified manner (\ref{old}). More precisely we replaced
\be \label{olda}
\frac{1}{1-\Pi(s)}\rightarrow \left(\frac{1}{1-\Pi(s)}-\sum_V\frac{3m_V\Gamma_{ll}\alpha^{-1}}{m^2_V-s+im\Gamma}\right)
\ee
in the muon production cross section formula. For leptonic decay  widths  $\Gamma_{ll}$ we take the experimentally determined values \cite{dzejpsaj2004,PDG}. The function $\Pi(s)$ in (\ref{olda}) is the same as before (in fact, the formula (\ref{olda}) is actually used for the experimental determination of leptonic widths, noting that the appropriate meson content should be removed from the vacuum polarization function in monte-carlo simulators).   
As we can see in Fig.\ref{figure2}, the presence of quarkonia affected the shape of resonance, however what is gain in the right shoulder  gets lost in the left one.
As in the previous case we comment the figure here in the text: now the thick solid 
line stands for the result with quarkonia included, the dashed line is the identical with the solid line in the previous figure , i.e. it represents the result  without heavy quarkonia included, the continuous errors were estimated as in the previous case. We do not re-adjust the error function, it counts with a fake -basically 100 percentage- error at $J/\psi $ energy.
The resulting curves exhibit the shape structure (in-)sensitivity to the relatively large changes in $R_h$ at that scales.
 To get rid of $c\bar{c}, b\bar{b} $ content as "background" contribution in $\Pi_h$,   we have also  checked that the inclusion  of perturbative loops does not change previous discussion.

Another systematics could appear due to two photon exchanges in $\sigma_h$ as well as  in leptonic cross sections.
These are hard to estimate without explicit calculation, however another type  two initial photons process $e^+e^-\rightarrow
\eta \gamma$ offers a rough  estimate 
\be \label{haf}
\delta\sigma_{h(\gamma\gamma)}=\frac{4\sigma( e^+e^-\rightarrow
\eta \gamma)}{\pi}\simeq 10nb
\ee
being thus negligible or comparable to statistical error  (note, $\pi$ in Eq.(\ref{haf}) is due to the loop and the factor $4$ counts the number of corrected relevant processes).   

 The two photon box diagram contribution in $\mu-$ pair cross section is known and  its interference term with dressed Born term does not contribute to total KLOE cross section, but to charge asymmetry only. Hence  we can conclude
the obtained discrepancy between Theory and experiments remains unexplained in a framework of dispersion relation for $\Pi_h$.
Conservative estimate gives us $0.14 nb$ total averaged error for Theory  leptonic cross section. 
Concerning experimental choice of energies by KLOE Collaboration in 2004, these three numbers are located five standard deviations in average from theoretically determined  values (with heavy quarkonia included in $\sigma_h$).

\section{Conclusion and perspectives}

In this paper we have shown that this millennium published data for electromagnetic productions of hadrons reached  necessary precision to test Standard Theory by study of  interference effect in the vacuum polarization function. We have calculated hadronic polarization function by using Optical Theorem  and used it for calculation of muon pair production.
 We have found a certain tension between  measured $\mu-$ pair cross section by KLOE and calculated one. The experimental/theory difference is nearly five percentage when the slope of interference structure is considered. The discrepancy alone does not represent a compelling signal for new physics beyond Standard Model and should be considered in a wider frame of other theoretical constrains and experimental signs.   

We did not use older data for leptonic cross sections. They can be analyzed in similar way  if they are generally available for public. Also similar analyzes can be applied to heavier quarkonia energy, which is recently  beyond the scope of presented paper. To the best knowledge of the author,  no discrepancy has been observed  or  announced via existing and working Monte-Carlo generators at
 $J/\Psi$ region.

\begin{figure}
\centerline{\epsfig{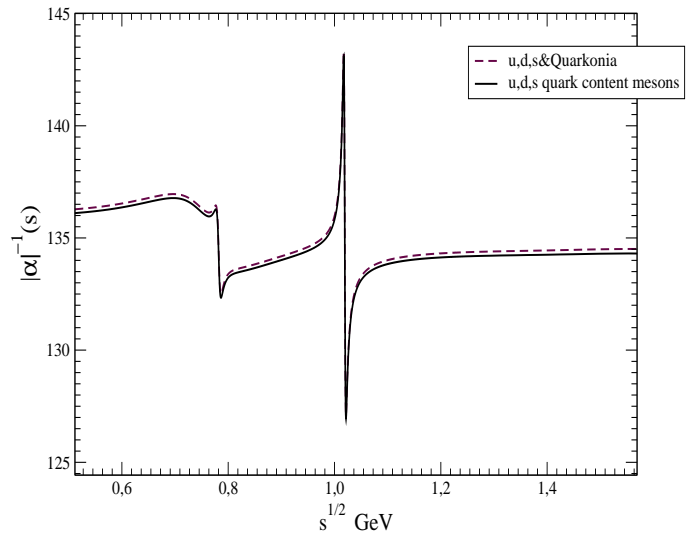}}
\caption[caption]{QED running coupling calculated  without and with heavy quarkonia included.}\label{running} 
\end{figure}

Within a new experiments underway, like CMD-3  SND2 and Belle-2 or working yet  BESIII, there is an easy  chance to improve theoretical predictions for leptonic cross section at $\phi$-meson energy region.   
As a bonus we get also similar interference effect at $\omega$ region. This somehow smaller, $3nb$ sized  zig-zag structure defined by its shoulders positions  in the window $0,778GeV ; 68nb$ - $0,0786GeV; 70,5 nb$ if virtual KLOE was operating there, has never been measured in an experiment precisely and is waiting for future generation of muonic detectors. In this special case, the precision of measured $\sigma_h$ allows to  compete classical QED experiments through the theory. The theory-experiment comparison in the 
$\rho/\omega$ region is remaining challenge for both the Standard Theory and new generation experiments as well. The way the $\rho/\omega$ and $\phi$ peaks are pronounced in the QED running  coupling is shown in Fig. \ref{running}. 

%%%%%%%%%%%%%%%%%%%%%%%%%%%%%%%%%%%%%%%%%%%%%%%%%%%%%%%%%%%%%%%%%%%%%%%

%
\end{document}